\documentclass[a4paper,11pt]{article}
\usepackage{array}
\usepackage{jinstpub} 

\title{\boldmath Calculation of true coincidence summing correction factor for clover detector in add-back and direct mode}

\author[a,b]{Ashish Gupta,}
\author[a,1]{M. Shareef,\note{Present address: Department of Physics, National Institute of Technology Calicut, Kozhikode 673601, India}}
\author[a,b]{Munmun Twisha,}
\author[a,b]{Saikat Bhattacharjee,}
\author[a,b,2]{A. Mukherjee,\note{Corresponding author.}}

\affiliation[a]{Saha Institute of Nuclear Physics,\\1/AF, Bidhan Nagar, Kolkata, 700064, India}
\affiliation[b]{Homi Bhabha National Institute,\\Anushaktinagar, Mumbai, 400094, India}

\emailAdd{anjali.mukherjee@saha.ac.in}

\abstract{The true coincidence summing effect on the full-energy peak efficiency calibration of an unsuppressed clover HPGe detector has been studied. Standard multi-energetic and mono-energetic $\gamma$-ray sources were used to determine the full-energy peak efficiency of the detector as a function of the $\gamma$-ray energies at different source-to-detector distances. The true coincidence summing correction factors for the full-energy peak efficiency of the detector has been determined, in the add-back and direct modes of the detector, using both experimental and analytical methods. Geant4 simulations were performed to obtain the full-energy peak efficiency and total efficiency of the detector for different $\gamma$-ray energies. The simulated efficiencies were used to calculate the correction factors using the analytical method. The correction factors obtained from both analytical and experimental methods were found to be in good agreement with each other. The clover detector in add-back mode exhibits larger summing corrections compared to the direct mode for the same source-to-detector distances. For the add-back mode, the coincidence summing effect is not significant for source-to-detector distances $\sim$ 13 cm or above, whereas, for the direct mode, measurements can be performed for source-to-detector distances $\sim$ 5 cm or above without considering the coincidence summing effect.}

\keywords{Gamma detectors (scintillators, CZT, HPGe, HgI etc); Detector modelling and simulations I (interaction of radiation with matter, interaction of photons with matter, interaction of hadrons with matter, etc)}

\arxivnumber{1234.56789} 

\proceeding{N$^{\text{th}}$ Workshop on X\\
  when\\
  where}

\begin{document}
\maketitle
\flushbottom

\section{Introduction}
\label{introduction}
In low energy nuclear physics and nuclear astrophysics experiments, one often needs to measure very small reaction cross-sections. Such measurements are considerably difficult because of the low rate of production of the reaction products. The rate of production can be boosted by increasing the beam current, the target thickness, and the irradiation time. But, all these parameters have practical limitations and so can not be increased infinitely. The other parameter that can help in increasing the statistics is the detector efficiency.

It is well known that as source-to-detector distance ($d$) decreases, the efficiency of the detector increases. The increased efficiency increases the detected yield because of the larger solid angle subtended by the detector compared to that for a distant geometry measurement. A close geometry measurement therefore helps in detecting the events with relatively higher statistics and hence determines the cross-section with better certainty. In $\gamma$-ray spectroscopy, High Purity Germanium (HPGe) detectors are often used because of their good energy resolution. In recent times, large volume single crystal HPGe detectors are often used because of their higher efficiencies compared to the conventional single crystal HPGe detector. With the large volume HPGe detector, one gets the efficiency enhancement factor of 2 at 100 keV and 3 at 1 MeV in detector efficiency \cite{bellia1989performances} compared to the conventional single crystal HPGe detector. However, the large-volume crystal has poor timing characteristics, which means it takes more time for the collection of electron-hole pair \cite{joshi1997study}. Apart from the difficulty in producing large volume crystals, ballistic deficit, neutron damage sensitivity, and large Doppler broadening problems are also encountered for such large volume single crystal HPGe detectors. By contrast, a clover detector consisting of four small single HPGe crystals inside a single cryostat may be a better choice. Each crystal of the clover detector behaves as a separate detector and output from each crystal is collected separately, thereby preserving its timing characteristics as well as a better time and energy resolution with higher efficiency \cite{sarkar2002characteristics, duchene1999clover}. The clover detector has approx 470 cc$^3$ of active volume \cite{agarwal2014coincidence} and presents good energy and timing resolutions for $\gamma$-ray full-energy peak (FEP) efficiency. The data from the four crystals in a clover detector can be collected separately and then added to generate the sum spectrum which is called the direct mode of data collection. On the other hand, the data from the four crystals can also be collected in a list mode and then generate an add-back spectrum.  In the add-back mode, the deposited energy in different crystals corresponding to a given event is summed in order to reconstruct the full-energy signal. The summed signal is stored in the add-back spectrum, which results in improved FEP efficiency. In the direct mode, each crystal is considered as a separate detector, and output from each crystal is added and the FEP efficiencies were calculated \cite{dababneh2004gamma}. In the add-back mode, the clover detector has a higher efficiency than the direct mode. However, for close-geometry measurements, though the clover detector in add-back mode may appear fine with its increased efficiency, it also introduces the coincidence summing effect, which is an unwanted effect. 

\begin{figure}
    \centering
    \includegraphics[scale = 0.5]{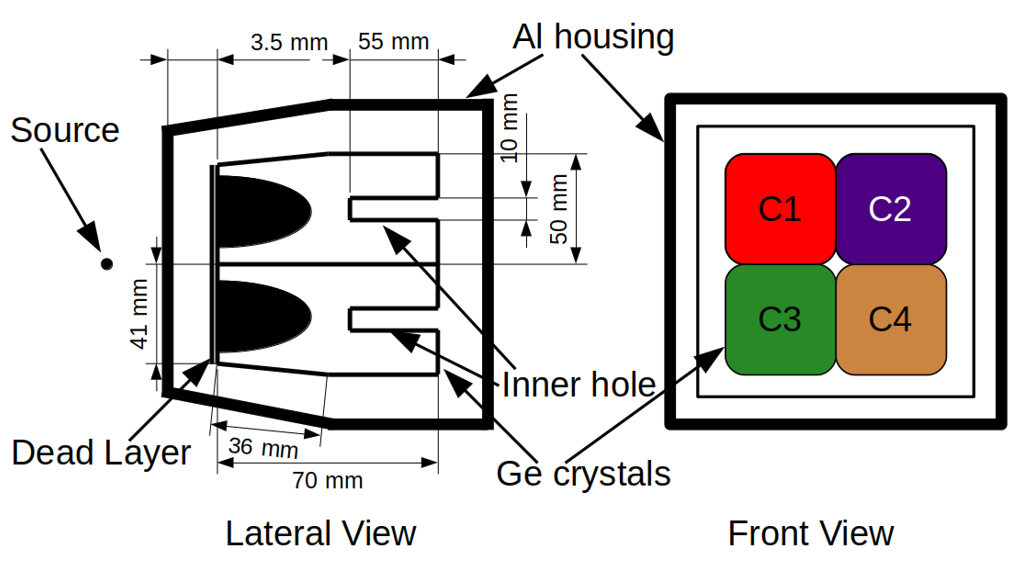}
    \caption{Schematic diagram of a clover detector with structural dimensions as provided by the manufacturer. C1, C2, C3, and C4 are four Germanium crystals of the clover detector. }
    \label{fig:clover_diagram}
\end{figure}

In $\gamma$-ray spectroscopy, true coincidence summing (TCS) occurs when two or more $\gamma$-rays emitted by a radioactive source in the cascade are detected within the detector's resolving time  \cite{gupta2023calculation, semkow1990coincidence}. The TCS effect is not only observed when two or more cascading $\gamma$-rays are detected simultaneously but also when a $\gamma$-ray and an X-ray are detected simultaneously \cite{mccallum1975influence}. The X-ray can be generated by the electron capture transition or by the internal conversion of the $\gamma$-ray in the decay scheme. The TCS effect depends on the source-to-detector distance ($d$) and the decay scheme of the $\gamma$-ray emitting source \cite{gupta2023calculation, debertin1979coincidence}. The radioactive source with a complex decay scheme will contribute more to the coincidence summing effect. The TCS correction is very important in close geometry measurements. The first order approximation for true coincidence summing correction factor ($k_{TCS}$) is given by \cite{agarwal2011true}, 

\begin{equation}\label{eq:analytical_correction_factor}
    k_{TCS} = \dfrac{1}{1-\Sigma_{i=1}^{i=n}p_i\epsilon_{ti}}
\end{equation}

 where n is the number of $\gamma$-rays which are in coincidence with the $\gamma$-ray of interest, $p_i$ is the probability of the simultaneous emission of the $i^{th}$ $\gamma$-ray to the $\gamma$-ray of interest and $\epsilon_{ti}$ is the total efficiency of the $i^{th}$ $\gamma$-ray. To calculate the correction factor using this analytical method requires the FEP and total efficiency of the $\gamma$-ray. This method has been described in references \cite{andreev1972consideration, debertin1988gamma, lin1991correction}. The information on the radioactive decay parameters of a source, such as the energies of the $\gamma$-ray transition, emission probabilities, decay modes of parent nuclei, mean energy of the K X-rays, total and K conversion coefficients, fluorescence yield, K-capture probabilities, are also required in this method to calculate the probability of the simultaneous emission of two or more $\gamma$-rays. The total efficiency is calculated using the spectra simulated by the Geant4 Monte Carlo simulation code \cite{geant4usermanual}. It is not possible to decompose the experimental spectra into well-defined components corresponding to $\gamma$-rays with distinct energies. Constructing the spectra for total efficiency using a multi-energetic source is therefore a very difficult task. The mono-energetic $\gamma$-ray sources have a single FEP, but the availability of such sources in the laboratories is very limited, and often the half-lives ($t_{1/2}$) of these sources are significantly small ($\sim$ days), and so we need to replace them periodically. Rather, one can simulate the detector response and generate the spectrum, and then find out the FEP and total efficiency from the simulated spectrum. The advantage of this method is that the simulated spectrum is free from the coincidence summing effect because the simulations are performed with mono-energetic $\gamma$-ray sources. The ratio of the efficiency obtained from the Geant4 simulation to the efficiency measured with multi-energetic $\gamma$-ray source is the experimental correction factor for the coincidence summing effect, which is given by,

 \begin{equation}\label{eq:exp_correction_factor}
     k_{TCS}^{Exp} = \dfrac{Geant4~~efficiency}{Experimental~~efficiency}
 \end{equation}

In our previous work \cite{gupta2023calculation}, TCS correction factors were calculated using experimental as well as analytical methods for a Falcon 5000, Broad Energy Germanium, single crystal, detector. In the present work, the TCS correction factors are determined for a clover detector, both in add-back mode and direct mode. 

The effect of coincidence summing on FEP efficiency for the add-back and direct mode of a clover detector at different source-to-detector distances ($d$ = 25, 13, 5, and 1 cm) has been studied. The $k_{TCS}$ for the FEP $\gamma$-rays from the radioactive sources, $^{60}$60, $^{133}$Ba and $^{152}$Eu have been calculated. The clover detector response has been simulated using the Geant4 simulation code. The mono-energetic $\gamma$-ray sources were used for optimizing the detector geometry to match the experimental efficiency with the simulated efficiency. The optimized detector geometry was later used to calculate the FEP and total efficiency of the $\gamma$-rays of present interest in order to calculate the coincidence summing correction factor for the clover detector, in add-back and direct modes. A comparison of correction factors in add-back and direct modes using analytical and experimental methods has been demonstrated in this work.

\begin{figure}
    \centering
    \includegraphics[trim={1.5cm 1.5cm 2cm 1cm},clip,scale=0.6]{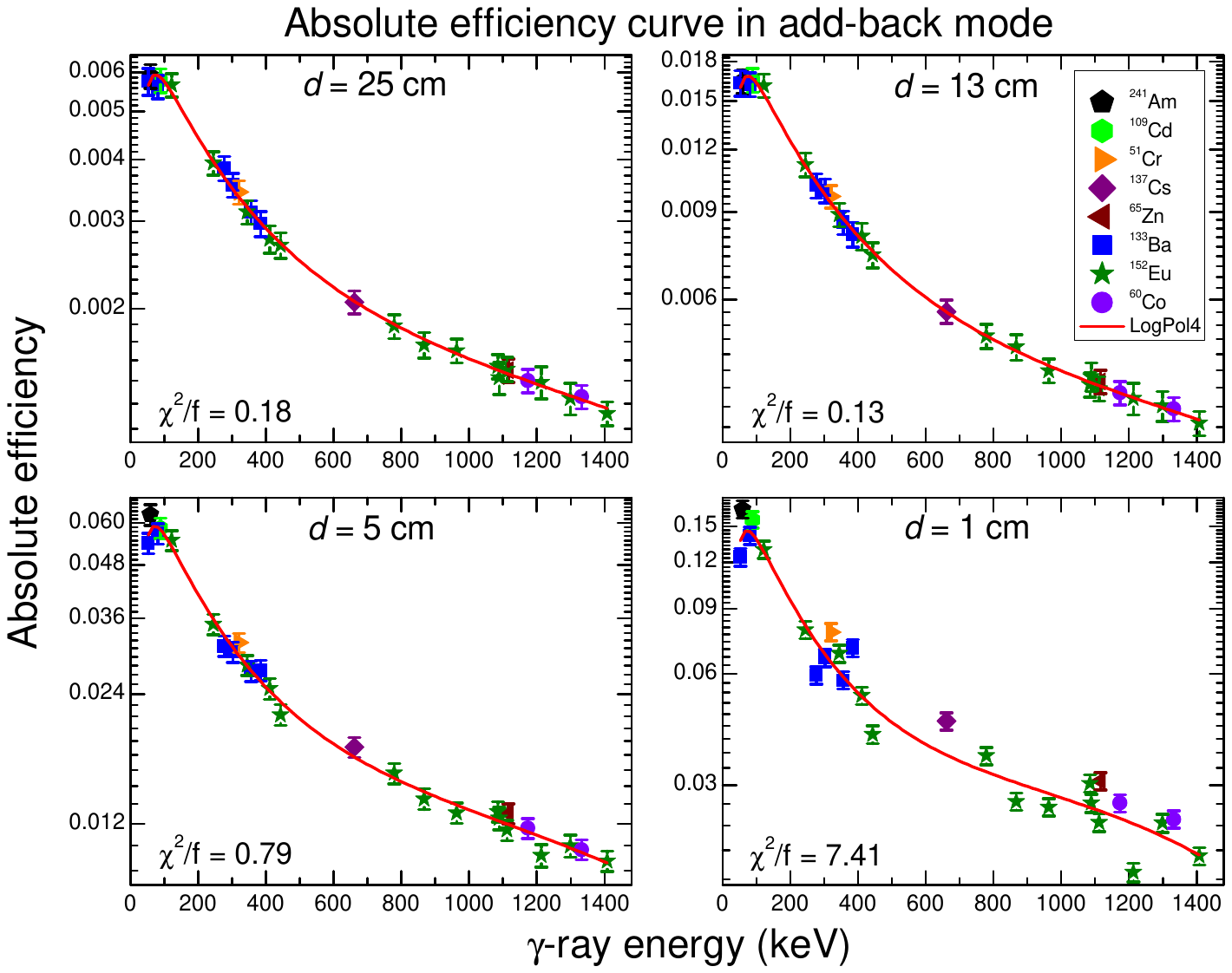}
    \caption{Experimental efficiency for the add-back mode of clover detector as a function of $\gamma$-ray energy at different source-to-detector distances, $d$ = 25, 13, 5 and 1 cm. The solid lines show the fitted curves using the $4^{th}$ order log-log polynomial.}
    \label{fig:exp_addback}
\end{figure}

\section{Experiment}
\label{sec:experiment}
To determine the absolute efficiency of an unsuppressed clover detector, calibrated radioactive sources, like $^{60}$60, $^{133}$Ba, and $^{152}$Eu were placed in front of the detector. The clover detector, used in this work was of CANBERRA make. The detector consists of 4 co-axial N-type Ge crystals, mounted in a single cryostat with one single end cap with a tapered rectangular section, each crystal has a diameter of 50 mm and a length of 70 mm. The distance between two adjacent Ge crystals was about 0.2 mm. The total active volume of the clover detector was 470 cm$^3$. A schematic diagram of the clover detector used in the study is shown in figure \ref{fig:clover_diagram} with its dimensional parameters. The energy resolution of each Ge crystal was less than 2 keV for the $\gamma$-rays at 1173, 1332 keV. The bias voltage of +2700 V was supplied to the clover detector by the ORTEC-660 high voltage power supply module. The output of each Ge crystal of the clover detector was fed as input to the CAEN desktop digitizer DT5725S module \cite{CaenDigitizer} which was controlled by the CoMPASS software \cite{CaenCopmpass}. The events with its timestamp from each crystal were recorded in a separate file. The add-back spectrum of the clover detector, generated by the software, was stored separately in an output file. The generated output files were stored in ROOT file format. The final offline data analysis was performed using the ROOT package \cite{brun1997root}. Initially, measurements were performed at a distance, $d$ = 25 cm from the face of the detector. This large distance was chosen to avoid the effect of coincidence summing in the measurements. Later, the radioactive sources were also placed in close geometry measurements for the distances $d$ = 13, 5, and 1 cm.

\begin{figure}
    \centering
    \includegraphics[trim={1.8cm 1.5cm 2cm 1cm},clip,scale=0.6]{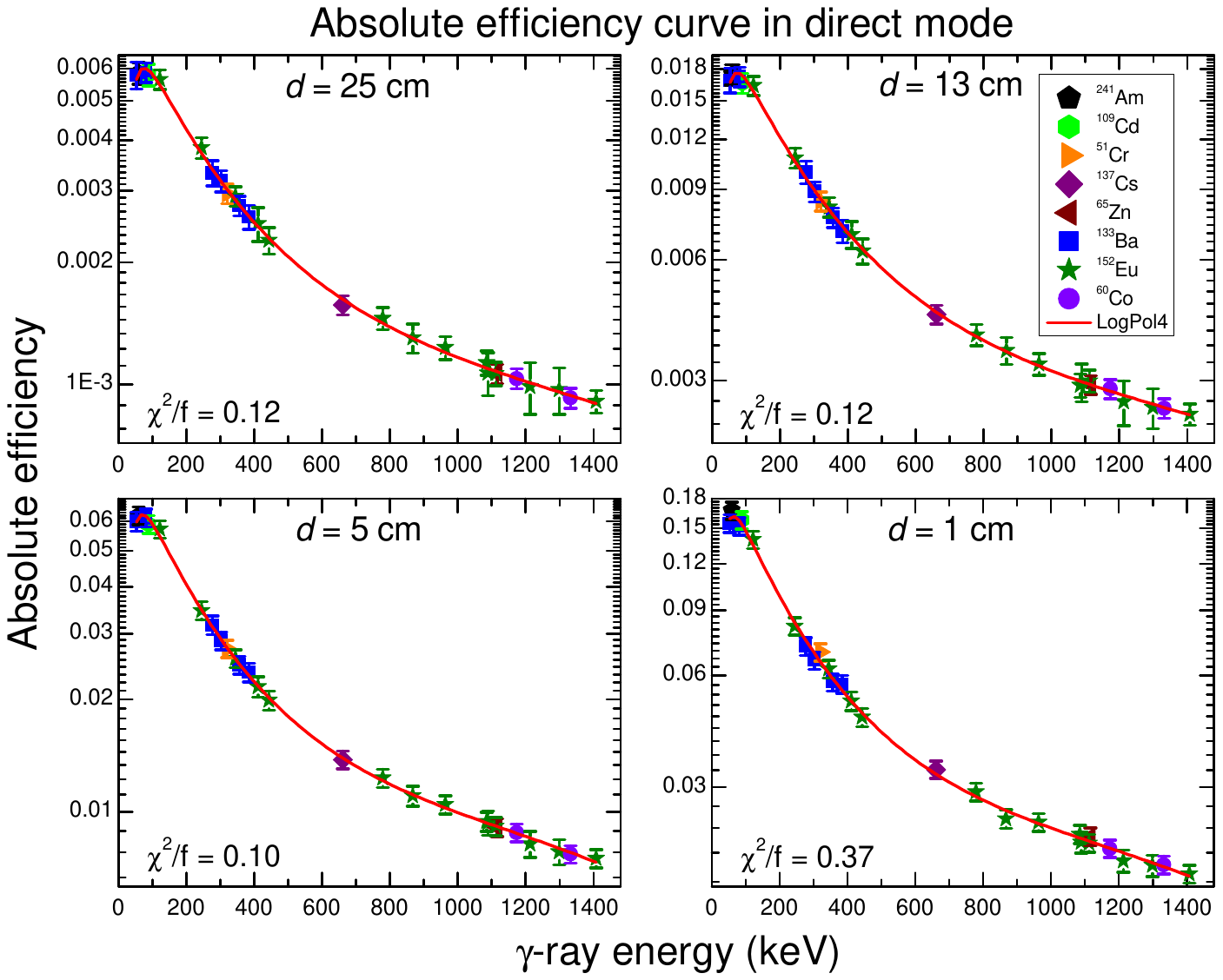}
    \caption{Experimental efficiency for the direct mode of clover detector as a function of $\gamma$-ray energy at different source-to-detector distances, $d$ = 25, 13, 5 and 1 cm. The solid lines show the fitted curves using the $4^{th}$ order log-log polynomial.}
    \label{fig:exp_sum}
\end{figure}

\begin{figure}
    \centering
    \includegraphics[scale=0.8]{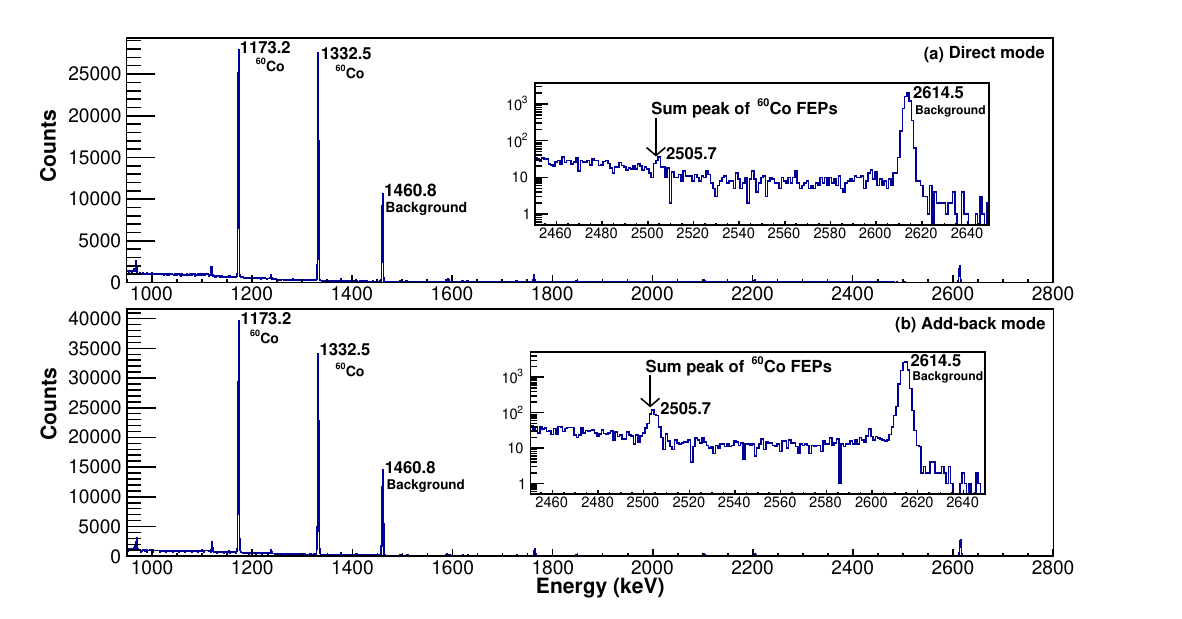}
    \caption{$\gamma$-ray spectrum of the clover detector with $^{60}$Co source at $d$ = 13 cm in (a) direct mode and (b) add-back mode.  In the inset of each panel, a zoomed view of the spectrum for the $\gamma$-ray energy region $\sim$ 2450-2650 keV is shown.}
    \label{fig:histogram_13cm}
\end{figure}

\begin{figure}
    \centering
    \includegraphics[scale=0.8]{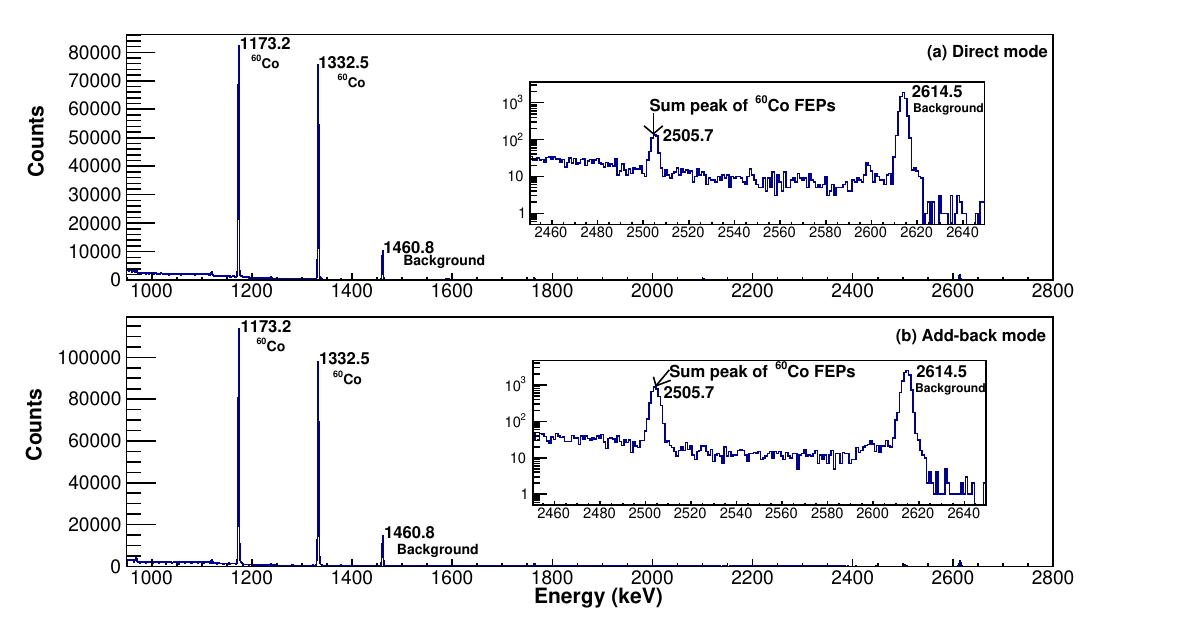}
    \caption{$\gamma$-ray spectrum of the clover detector with $^{60}$Co source at $d$ = 5 cm in (a) direct mode and (b) add-back mode. In the inset of each panel, a zoomed view of the spectrum for the $\gamma$-ray energy region $\sim$ 2450-2650 keV is shown.}
    \label{fig:histogram_5cm}
\end{figure}

The clover detector was operated in both, add-back and direct modes. The FEP efficiencies for the clover detector in different modes were calculated using the mono-energetic $\gamma$-ray sources $^{51}$Cr, $^{65}$Zn, $^{109}$Cd, $^{137}$Cs, and $^{241}$Am and multi-energetic $\gamma$-ray sources $^{60}$Co, $^{133}$Ba, and $^{152}$Eu. The radioactive sources $^{60}$Co, $^{133}$Ba, $^{137}$Cs, $^{152}$Eu and $^{241}$Am were available in the laboratory and the sources  $^{51}$Cr, $^{65}$Zn and $^{109}$Cd were prepared \cite{gupta2023calculation} by irradiating foils of $^{51}$V, $^{65}$Cu and $^{109}$Ag with proton beam from the K-130 cyclotron at VECC, Kolkata \cite{bhattacharya2018experimental}. The measured FEP efficiencies of the clover detector for add-back mode as a function of the $\gamma$-ray energy are shown in figure \ref{fig:exp_addback}. The efficiency data points at each distance $d$, were fitted with a $4^{th}$ order log-log polynomial and are shown in the figures by the solid lines. The value of $\chi^2/f$ (chi-square per degree of freedom) given in each of the figures shows the quality of fit. From the figure \ref{fig:exp_addback}, it is clearly seen that at distances $d$ = 25 and 13 cm efficiency data points match very well with the fitted curves. Whereas, at distances $d$ = 5 and 1 cm, the efficiency data points deviate from the fitting curves, with deviation increasing with a decrease of $d$. Figure \ref{fig:exp_sum} shows the measured FEP efficiencies of the clover detector for direct mode as a function of the $\gamma$-ray energies. In the direct mode, the efficiency data points match well with the fitted curves at distances $d \ge $ 5 cm. Whereas, at distance $d$ = 1 cm data points deviate from the fit curve. The deviation from the fit curve of the experimentally determined efficiency of the clover detector in add-back and direct mode is due to the presence of coincidence summing effect in close geometry. The $\gamma$-ray spectrum of the clover detector with $^{60}$Co source at $d$ = 13 cm has been shown in figure \ref{fig:histogram_13cm}. In the inset of figure \ref{fig:histogram_13cm}(a), even though one can see every small sum peak at 2505.7 keV, the individual contribution from the full energy peaks, 1173.2 keV and 1332.5 keV towards the sum peak will be almost half of the area under the curve at 2505.7 keV peak. Hence the loss of FEP towards the sum peak in the direct mode is almost insignificant. However, as seen from the inset of figure \ref{fig:histogram_13cm}(b) in the add-back mode, there is a considerable contribution of the sum peak. This shows for the add-back mode, the coincidence summing effect becomes significant at $d \sim$  13 cm. The $\gamma$-ray spectrum of the clover detector with $^{60}$Co source at $d$ = 5 cm has been shown in figure \ref{fig:histogram_5cm}. A sum peak in the add-back mode can be clearly seen whereas in the direct mode, there is a small presence of a sum peak but it has been found that this does not have any significant effect on the direct mode $\gamma$-ray efficiency. The presence of the sum peak at $d$ = 13 cm in add-back mode and at $d$ = 5 cm in direct mode confirms the importance of the coincidence summing effect in close geometry in the respective modes.

\begin{figure}
    \centering
    \includegraphics[trim={1cm 2cm 1.5cm 1.7cm},clip,scale = 0.6]{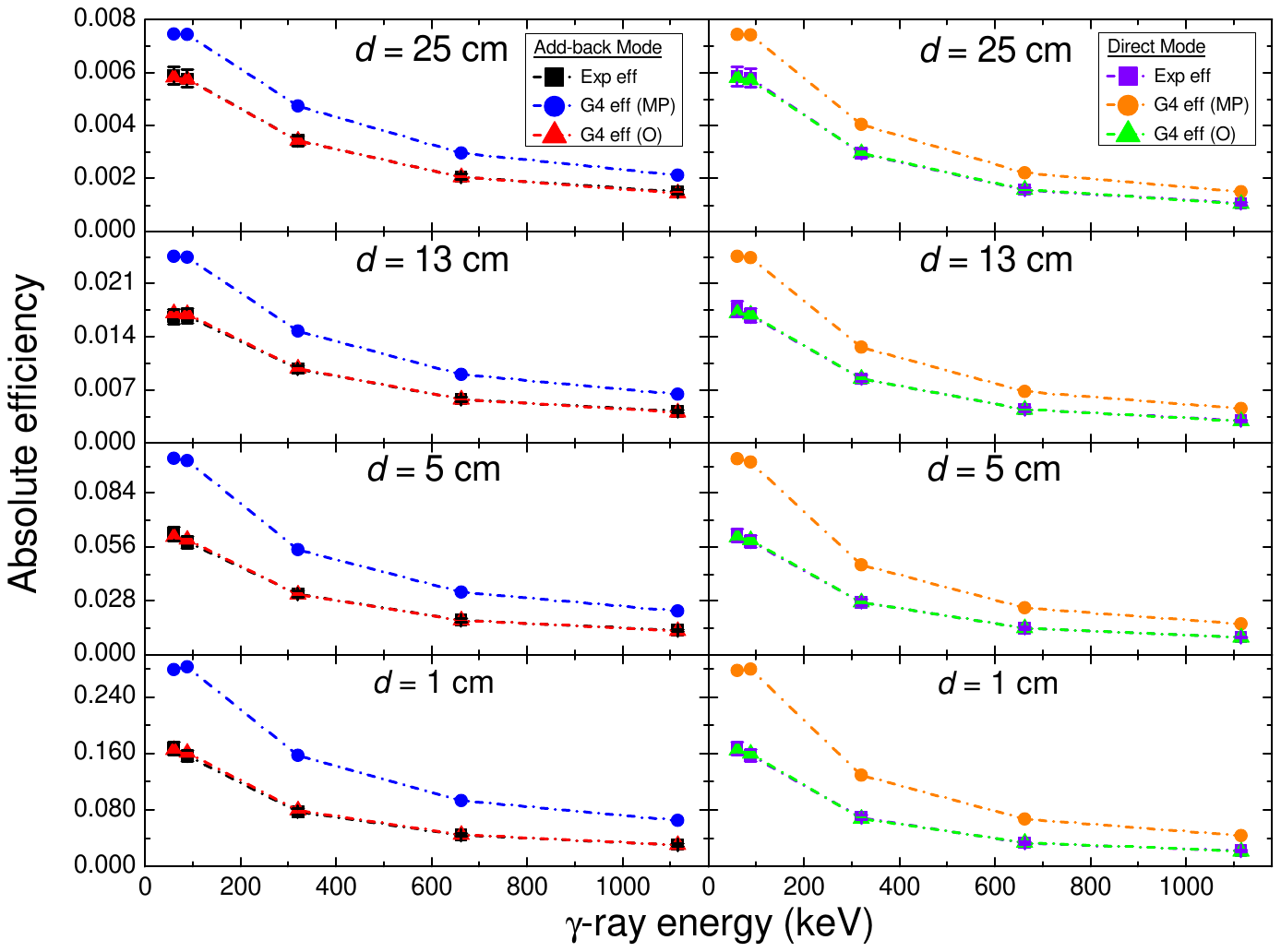}
    \caption{Efficiencies of mono-energetic $\gamma$-ray sources at $d$ = 25, 13, 5 and 1 cm. Experimental efficiencies are shown by the solid squares. The solid circles show the Geant4 (G4) simulated efficiencies with the detector parameter provided by the manufacturer (MP) and the solid triangles show the Geant4 simulated efficiencies with the optimized (O) detector parameter set. The left panel data points correspond to the add-back mode of the clover detector and the right panel data correspond to the direct mode of the clover detector. Lines are drawn to guide the eye.}
    \label{fig:monoenergetic}
\end{figure}

One way to avoid the coincidence summing effect in close geometry measurements is to use the mono-energetic $\gamma$-rays sources. But the availability of these sources for a wide energy range is usually not readily available, as already mentioned in section \ref{introduction}. So it is a standard practice to use the multi-energetic $\gamma$-ray sources to determine the efficiency of the detector, and then for close geometry measurements, one has to do the summing correction. To deal with the summing effect, TCS correction has to be done for detector efficiency measurements. The TCS correction factor can be calculated by experimental and analytical methods. To calculate the correction factor with these methods, we need the probability of the coincidence of $\gamma$-rays, FEP, and total efficiency of corresponding $\gamma$-rays. In this work, the detector response at different distances $d$ was simulated using the Geant4 simulation toolkit \cite{agostinelli2003geant4, pia2003geant4}.

\begin{table}[]
    \centering
    \caption{Detector dimension parameters provided by the manufacturer and optimized with Geant4 simulation.}
    \begin{tabular}{m{4cm} m{4cm} m{3.6cm} }
    \hline
    Detector parameters &   Manufacturer dimensions (mm) & Optimized dimensions (mm)\\
    \hline
     Crystal radius         &       25.00    &  23.20   \\
     Crystal length         &       70.00    &  70.00   \\
     Dead layer             &       0.0005   &  0.0005  \\
     Inner hole radius      &       5.00     &  5.50    \\
     Inner hole depth       &       55.00    &  55.00   \\
     Al end cap thickness   &       1.50     &  1.00    \\
     Al end cap to crystal distance & 3.50   & 21.30    \\
     \hline
    \end{tabular}
    \label{tab:Detector parameters}
\end{table}

\begin{figure}
    \centering
    \includegraphics[trim={0.5cm 0.5cm 2cm 2cm},clip,scale=0.6]{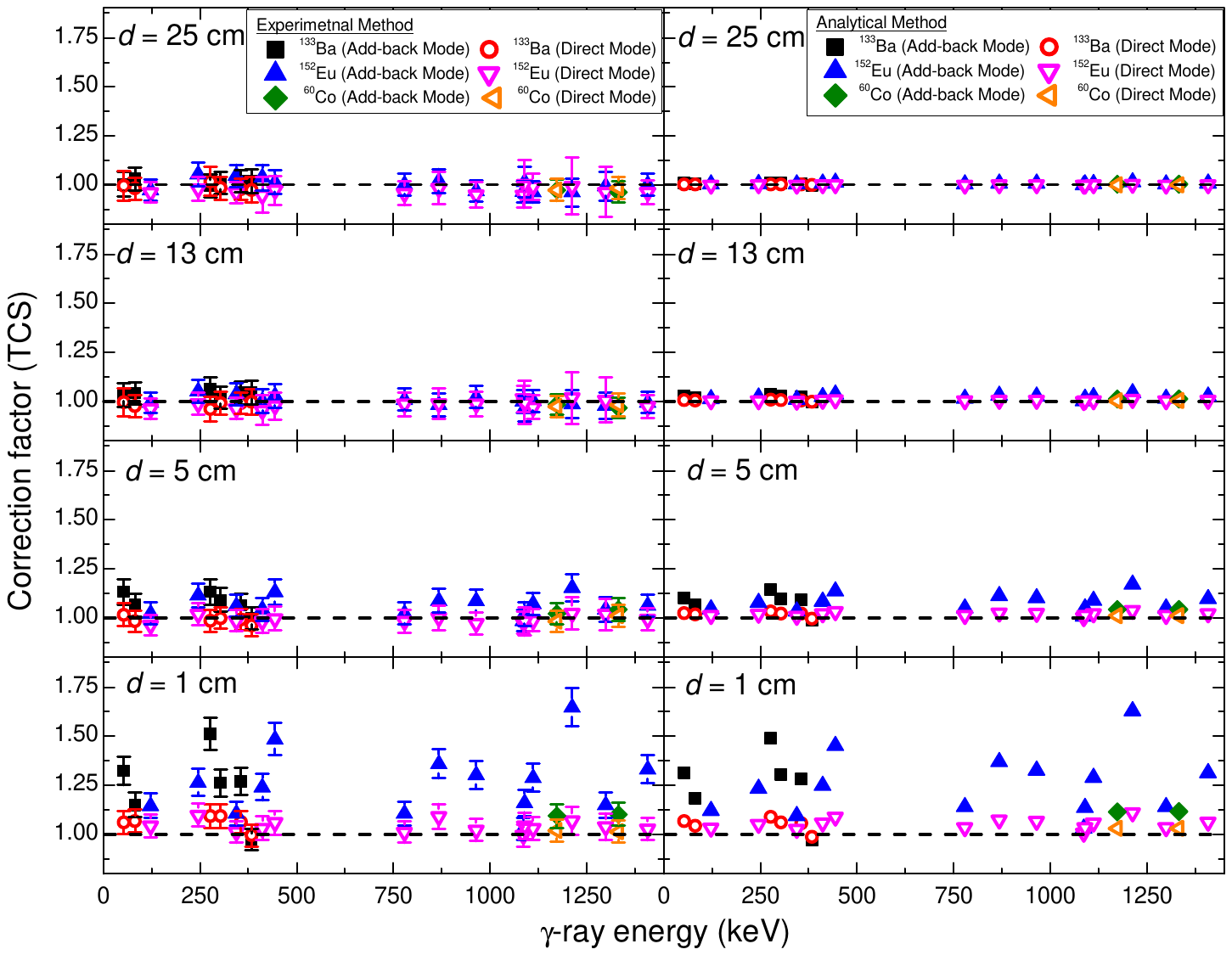}
    \caption{Comparison of correction factors (TCS) for the add-back and direct mode of clover detector using experimental method (left panel) as well as analytical method (right panel) at distances, $d$ = 25, 13, 5 and 1 cm.}
    \label{fig:correction_factor}
\end{figure}

\begin{table}[]
    \centering
    \caption{Comparison of the experimental efficiencies (Exp eff) for the $\gamma$-rays from mono-energetic $\gamma$-ray sources with the Geant4 efficiencies (G4 eff) obtained from Geant4 simulation in add-back and direct mode at different distances $d$ = 25, 13, 5 and 1 cm. The simulation was done using manufacturer-provided (MP) and optimized (O) parameter sets. }
    \resizebox{\textwidth}{!}{
    \begin{tabular}{l ||p{2cm} p{1.1cm} p{1.3cm} p{1.1cm} p{1.3cm} ||p{2cm} p{1.1cm} p{1.3cm} p{1.1cm} p{1.3cm}}
\hline
E$_\gamma$ (keV) & Exp eff & G4 eff (MP) & G4/Exp (MP) & G4 eff (O) & G4/Exp (O) &  Exp eff & G4 eff (MP) & G4/Exp (MP) & G4 eff (O) & G4/Exp (O)\\
\hline																					
\underline{$d$ = 25 cm}	&\multicolumn{5}{c||}{\underline{Add-back mode}}  & \multicolumn{5}{c}{\underline{Direct mode}}   \\
59.5	&	0.00589 (33)	&	0.0075	&	1.26	&	0.0058	&	0.99	&	0.00585 (36)	&	0.0074	&	1.27	&	0.0058	&	0.99	\\
88	&	0.00577 (32)	&	0.0075	&	1.29	&	0.0057	&	0.99	&	0.00579 (35)	&	0.0074	&	1.28	&	0.0057	&	0.99	\\
320.1	&	0.00343 (19)	&	0.0047	&	1.38	&	0.0034	&	1.00	&	0.00295 (16)	&	0.0040	&	1.37	&	0.0030	&	1.00	\\
661.6	&	0.00206 (11)	&	0.0030	&	1.43	&	0.0021	&	1.00	&	0.00157 (09)	&	0.0022	&	1.42	&	0.0016	&	1.01	\\
1115.5	&	0.00150 (08)	&	0.0021	&	1.42	&	0.0015	&	0.97	&	0.00106 (06)	&	0.0015	&	1.42	&	0.0011	&	0.99	\\
Avg. ratio 	&      	            	&	    	&	1.36	&	    	&	0.99	&	    	&	    	&	1.35	&	    	&	1.00	\\ 
std dev	&		&		&	0.07	&		&	0.01	&		&		&	0.06	&		&	0.01	\\
\% err	&      		&      		&      	4.99	&      		&      	1.23	&      		&      		&      	4.67	&      		&      	0.81	\\ 
\hline																					
\underline{$d$ = 13 cm}	&		&		&		&		&		&		&		&		&		&		\\
59.5	&	0.0164 (09)	&	0.0243	&	1.48	&	0.0169	&	1.03	&	0.0175 (10)	&	0.0242	&	1.39	&	0.0169	&	0.97	\\
88	&	0.0165 (09)	&	0.0242	&	1.46	&	0.0168	&	1.02	&	0.0166 (10)	&	0.0241	&	1.45	&	0.0167	&	1.01	\\
320.1	&	0.0097 (05)	&	0.0145	&	1.51	&	0.0097	&	1.01	&	0.0084 (05)	&	0.0125	&	1.48	&	0.0084	&	1.00	\\
661.6	&	0.0057 (03)	&	0.0089	&	1.57	&	0.0057	&	1.00	&	0.0044 (02)	&	0.0067	&	1.53	&	0.0044	&	1.00	\\
1115.5	&	0.0041 (02)	&	0.0063	&	1.54	&	0.0040	&	0.97	&	0.0029 (02)	&	0.0045	&	1.54	&	0.0029	&	0.99	\\
Avg. ratio 	&      	            	&	    	&	1.51	&	    	&	1.01	&	    	&	    	&	1.48	&	    	&	0.99	\\ 
std dev	&		&		&	0.04	&		&	0.02	&		&		&	0.06	&		&	0.01	\\
\% err	&      		&      		&      	2.55	&      		&      	1.93	&      		&      		&      	3.76	&      		&      	1.31	\\ 
\hline																					
\underline{$d$ = 5 cm}	&		&		&		&		&		&		&		&		&		&		\\
59.5	&	0.0628 (34)	&	0.1021	&	1.63	&	0.0613	&	0.98	&	0.0621 (34)	&	0.1019	&	1.64	&	0.0612	&	0.99	\\
88	&	0.0584 (32)	&	0.1008	&	1.73	&	0.0599	&	1.03	&	0.0589 (32)	&	0.1003	&	1.70	&	0.0596	&	1.01	\\
320.1	&	0.0316 (17)	&	0.0546	&	1.73	&	0.0315	&	1.00	&	0.0273 (15)	&	0.0467	&	1.71	&	0.0272	&	1.00	\\
661.6	&	0.0181 (10)	&	0.0325	&	1.80	&	0.0180	&	1.00	&	0.0138 (07)	&	0.0244	&	1.77	&	0.0138	&	1.00	\\
1115.5	&	0.0127 (07)	&	0.0228	&	1.80	&	0.0124	&	0.98	&	0.0091 (05)	&	0.0161	&	1.78	&	0.0090	&	0.99	\\
Avg. ratio 	&      	            	&	    	&	1.74	&	    	&	0.99	&	    	&	    	&	1.72	&	    	&	1.00	\\ 
std dev	&		&		&	0.06	&		&	0.02	&		&		&	0.05	&		&	0.01	\\
\% err	&      		&      		&      	3.68	&      		&      	1.75	&      		&      		&      	2.91	&      		&      	0.92	\\ 
\hline																					
\underline{$d$ = 1 cm}	&		&		&		&		&		&		&		&		&		&		\\
59.5	&	0.1669 (90)	&	0.2789	&	1.67	&	0.1652	&	0.99	&	0.1673 (91)	&	0.2781	&	1.66	&	0.1649	&	0.99	\\
88	&	0.1566 (84)	&	0.2830	&	1.81	&	0.1611	&	1.03	&	0.1576 (85)	&	0.2800	&	1.78	&	0.1600	&	1.01	\\
320.1	&	0.0778 (42)	&	0.1573	&	2.02	&	0.0799	&	1.03	&	0.0694 (37)	&	0.1295	&	1.87	&	0.0682	&	0.98	\\
661.6	&	0.0447 (24)	&	0.0934	&	2.09	&	0.0449	&	1.01	&	0.0335 (18)	&	0.0671	&	2.00	&	0.0340	&	1.01	\\
1115.5	&	0.0308 (17)	&	0.0653	&	2.12	&	0.0305	&	0.99	&	0.0221 (12)	&	0.0441	&	1.99	&	0.0218	&	0.99	\\
Avg. ratio 	&      	            	&	    	&	1.94	&	    	&	1.01	&	    	&	    	&	1.86	&	    	&	1.00	\\ 
std dev	&		&		&	0.17	&		&	0.02	&		&		&	0.13	&		&	0.01	\\
\% err	&      		&      		&      	9.00	&      		&      	1.69	&      		&      		&      	6.98	&      		&      	1.49	\\

\hline

    \end{tabular}
    }
         \label{tab:monoenergetic_optimization}
\end{table}

\section{Geant4 Monte Carlo simulation}
\label{Geant4 Monte Carlo simulation}
The Geant4 toolkit simulates the passing and interaction of different particles through different media. Geant4 has been extensively used for the simulation of the efficiencies of different $\gamma$-ray detectors in the past \cite{dababneh2004gamma, sharma2020characterization, laborie2000monte}. To simulate the detector response, the geometry of the clover detector was reproduced by constructing four Germanium crystals of cylindrical shape with a hole of 55 mm depth in each, and tapered edges of the crystals were reproduced by subtracting a wedge shape from each side of each crystal \cite{saha2016geant4, palit2000parity}. Each of the crystals in the detector volume was placed so precisely, that they are as close as possible but do not overlap. After the construction of the detector geometry, the elements of the crystal were incorporated from the NIST library class which was inbuilt in Geant4. A 10$^8$ $\gamma$-rays were generated randomly and isotropically in all the directions in this simulation. The primary particle in the simulation, that is, the $\gamma$-rays interact with the matter through the electromagnetic process \cite{sharma2020characterization}. All the electromagnetic processes were included in the simulation using the inbuilt G4EmStandardPhysics class \cite{geant4usermanual}. The particles were monitored throughout the detector volume for each event. The deposited energy in germanium crystals was collected at each step and added at the end of each event, the total deposited energy in the detector was obtained by adding the energy deposited in the four crystals. These deposited energies for each crystal and the total deposited energy in the detector for each event were stored in ROOT file format at the end of each event and further offline analysis was done in the ROOT framework \cite{brun1997root}. 

\begin{table}[]
    \centering
     \caption{Coincidence summing correction factors using analytical ($k^{Analytical}_{TCS}$) and experimental ($k^{Exp}_{TCS}$) method for add-back and direct mode at distances $d$ = 25, 13,  5 and 1 cm. }
     \resizebox{\textwidth}{!}{
    \begin{tabular}{l m{1cm} || p{1.5cm} p{1cm} p{1.2cm} |  p{1.5cm} p{1cm} p{1.2cm} ||  p{1.5cm} p{1cm} p{1.2cm} | p{1.5cm} p{1cm} p{1.2cm}}
    
   \hline
    Source & E$_\gamma$ (keV) & $k^{Analytical}_{TCS}$ & $k^{Exp}_{TCS}$ & error in $k^{Exp}_{TCS}$ &  $k^{Analytical}_{TCS}$ & $k^{Exp}_{TCS}$ & error in $k^{Exp}_{TCS}$ & $k^{Analytical}_{TCS}$ & $k^{Exp}_{TCS}$ & error in $k^{Exp}_{TCS}$ & $k^{Analytical}_{TCS}$ & $k^{Exp}_{TCS}$ & error in $k^{Exp}_{TCS}$ \\
    \hline
         &          		& 	\multicolumn{3}{c|}{\underline{Add-back mode}}	&	\multicolumn{3}{c||}{\underline{Direct mode}}	   & \multicolumn{3}{c|}{\underline{Add-back mode}}  & \multicolumn{3}{c}{\underline{Direct mode}}    \\
    \hline
		 &          		& 	\underline{$d$ = 25 cm}	&		&		& 		&		&		& 	\underline{$d$ = 13 cm}	&		&		& 		&		&		\\
    $^{60}$Co	&	1173.2	&	1.00	&	0.97	&	0.05	&	1.00	&	0.98	&	0.06	&	1.01	&	0.98	&	0.05	&	1.00	&	0.97	&	0.05	\\
	&	1332.5	&	1.00	&	0.96	&	0.05	&	1.00	&	0.99	&	0.06	&	1.01	&	0.97	&	0.05	&	1.00	&	0.98	&	0.06	\\
    $^{133}$Ba	&	53.2	&	1.01	&	1.00	&	0.06	&	1.00	&	1.00	&	0.08	&	1.03	&	1.06	&	0.06	&	1.01	&	1.00	&	0.07	\\
	&	81	&	1.01	&	1.03	&	0.06	&	1.00	&	0.98	&	0.05	&	1.02	&	1.04	&	0.06	&	1.01	&	0.98	&	0.05	\\
	&	276.4	&	1.01	&	1.00	&	0.06	&	1.00	&	1.02	&	0.07	&	1.04	&	1.06	&	0.06	&	1.01	&	0.96	&	0.06	\\
	&	302.9	&	1.01	&	1.01	&	0.06	&	1.00	&	0.99	&	0.06	&	1.03	&	1.02	&	0.06	&	1.01	&	0.99	&	0.06	\\
	&	356	&	1.01	&	1.02	&	0.06	&	1.00	&	0.98	&	0.05	&	1.02	&	1.05	&	0.06	&	1.01	&	0.99	&	0.05	\\
	&	383.9	&	1.00	&	1.02	&	0.06	&	1.00	&	0.98	&	0.07	&	1.00	&	1.04	&	0.06	&	1.00	&	1.00	&	0.06	\\
    $^{152}$Eu	&	121.8	&	1.00	&	0.97	&	0.05	&	1.00	&	0.96	&	0.05	&	1.01	&	0.99	&	0.05	&	1.00	&	0.96	&	0.05	\\
	&	244.7	&	1.01	&	0.96	&	0.05	&	1.00	&	0.98	&	0.06	&	1.02	&	1.05	&	0.06	&	1.01	&	0.99	&	0.06	\\
	&	344.3	&	1.00	&	1.05	&	0.06	&	1.00	&	0.96	&	0.05	&	1.01	&	1.04	&	0.06	&	1.00	&	0.96	&	0.05	\\
	&	411.1	&	1.01	&	1.04	&	0.07	&	1.00	&	0.95	&	0.09	&	1.02	&	1.02	&	0.06	&	1.01	&	0.96	&	0.07	\\
	&	444	&	1.01	&	1.01	&	0.06	&	1.00	&	0.97	&	0.07	&	1.04	&	1.03	&	0.06	&	1.01	&	0.98	&	0.07	\\
	&	778.9	&	1.01	&	1.00	&	0.06	&	1.00	&	0.96	&	0.06	&	1.01	&	1.01	&	0.06	&	1.00	&	0.99	&	0.06	\\
	&	867.4	&	1.01	&	1.02	&	0.06	&	1.00	&	0.98	&	0.08	&	1.03	&	0.98	&	0.06	&	1.01	&	0.99	&	0.08	\\
	&	964.1	&	1.01	&	0.97	&	0.05	&	1.00	&	0.95	&	0.06	&	1.03	&	1.02	&	0.06	&	1.01	&	0.99	&	0.06	\\
	&	1085.9	&	1.00	&	0.97	&	0.05	&	1.00	&	0.95	&	0.07	&	1.00	&	1.00	&	0.06	&	1.00	&	1.01	&	0.07	\\
	&	1089.7	&	1.01	&	1.02	&	0.08	&	1.00	&	1.01	&	0.12	&	1.01	&	0.96	&	0.06	&	1.00	&	1.00	&	0.11	\\
	&	1112.1	&	1.01	&	0.96	&	0.05	&	1.00	&	0.99	&	0.07	&	1.02	&	1.00	&	0.06	&	1.01	&	0.97	&	0.11	\\
	&	1212.9	&	1.01	&	0.96	&	0.07	&	1.00	&	0.99	&	0.15	&	1.04	&	0.99	&	0.07	&	1.01	&	1.02	&	0.13	\\
	&	1299.1	&	1.01	&	0.99	&	0.07	&	1.00	&	0.96	&	0.13	&	1.01	&	0.98	&	0.07	&	1.00	&	1.01	&	0.12	\\
	&	1408	&	1.01	&	1.00	&	0.06	&	1.00	&	0.96	&	0.06	&	1.03	&	1.00	&	0.06	&	1.01	&	0.98	&	0.06	\\
\hline																											
    	 &          		& 	 \underline{$d$ = 5 cm}  	&		&		&		&           		& 		&	\underline{$d$ = 1 cm}            	&		& 		&		&		& 		\\
    $^{60}$Co	&	1173.2	&	1.04	&	1.02	&	0.06	&	1.01	&	0.98	&	0.05	&	1.11	&	1.10	&	0.06	&	1.03	&	1.02	&	0.06	\\
	&	1332.5	&	1.04	&	1.04	&	0.06	&	1.01	&	1.01	&	0.06	&	1.12	&	1.10	&	0.06	&	1.03	&	1.01	&	0.06	\\
    $^{133}$Ba	&	53.2	&	1.10	&	1.13	&	0.06	&	1.03	&	1.02	&	0.06	&	1.31	&	1.32	&	0.07	&	1.07	&	1.06	&	0.06	\\
	&	81	&	1.06	&	1.07	&	0.06	&	1.02	&	0.98	&	0.05	&	1.18	&	1.15	&	0.06	&	1.04	&	1.07	&	0.06	\\
	&	276.4	&	1.14	&	1.13	&	0.06	&	1.03	&	0.99	&	0.06	&	1.49	&	1.51	&	0.08	&	1.09	&	1.09	&	0.06	\\
	&	302.9	&	1.10	&	1.09	&	0.06	&	1.02	&	1.00	&	0.06	&	1.30	&	1.26	&	0.07	&	1.06	&	1.09	&	0.06	\\
	&	356	&	1.09	&	1.07	&	0.06	&	1.02	&	0.99	&	0.05	&	1.28	&	1.27	&	0.07	&	1.06	&	1.06	&	0.06	\\
	&	383.9	&	0.99	&	1.00	&	0.05	&	1.00	&	0.96	&	0.06	&	0.97	&	0.97	&	0.05	&	0.99	&	1.00	&	0.06	\\
    $^{152}$Eu	&	121.8	&	1.05	&	1.02	&	0.06	&	1.01	&	0.96	&	0.05	&	1.12	&	1.15	&	0.06	&	1.03	&	1.04	&	0.06	\\
	&	244.7	&	1.08	&	1.11	&	0.06	&	1.02	&	1.02	&	0.06	&	1.23	&	1.26	&	0.07	&	1.05	&	1.10	&	0.06	\\
	&	344.3	&	1.04	&	1.06	&	0.06	&	1.01	&	0.99	&	0.05	&	1.10	&	1.11	&	0.06	&	1.03	&	1.01	&	0.05	\\
	&	411.1	&	1.08	&	1.04	&	0.06	&	1.02	&	0.98	&	0.06	&	1.25	&	1.24	&	0.07	&	1.06	&	1.03	&	0.06	\\
	&	444	&	1.14	&	1.13	&	0.06	&	1.03	&	1.00	&	0.06	&	1.45	&	1.48	&	0.08	&	1.09	&	1.06	&	0.06	\\
	&	778.9	&	1.05	&	1.02	&	0.06	&	1.01	&	0.98	&	0.06	&	1.14	&	1.11	&	0.06	&	1.04	&	1.01	&	0.06	\\
	&	867.4	&	1.11	&	1.09	&	0.06	&	1.03	&	1.00	&	0.06	&	1.37	&	1.36	&	0.07	&	1.07	&	1.09	&	0.06	\\
	&	964.1	&	1.10	&	1.09	&	0.06	&	1.02	&	0.97	&	0.05	&	1.33	&	1.30	&	0.07	&	1.07	&	1.02	&	0.06	\\
	&	1085.9	&	1.01	&	0.99	&	0.05	&	1.00	&	0.98	&	0.06	&	1.03	&	1.03	&	0.06	&	1.01	&	1.00	&	0.06	\\
	&	1089.7	&	1.05	&	1.02	&	0.06	&	1.01	&	0.98	&	0.07	&	1.14	&	1.16	&	0.07	&	1.04	&	1.04	&	0.07	\\
	&	1112.1	&	1.09	&	1.07	&	0.06	&	1.02	&	0.99	&	0.06	&	1.29	&	1.29	&	0.07	&	1.06	&	1.03	&	0.06	\\
	&	1212.9	&	1.17	&	1.15	&	0.07	&	1.04	&	1.03	&	0.08	&	1.63	&	1.64	&	0.10	&	1.11	&	1.07	&	0.07	\\
	&	1299.1	&	1.05	&	1.04	&	0.06	&	1.01	&	1.02	&	0.08	&	1.14	&	1.15	&	0.06	&	1.04	&	1.04	&	0.07	\\
	&	1408	&	1.10	&	1.06	&	0.06	&	1.02	&	0.99	&	0.06	&	1.31	&	1.33	&	0.07	&	1.06	&	1.03	&	0.06	\\

   \hline
   \end{tabular}
   }
    \label{tab:correctionfactor}
\end{table}

Following the above procedure the $\gamma$-ray spectra for distances $d$ = 25, 13, 5, and 1 cm were simulated for add-back and direct mode, using the detector dimensions provided by the manufacturer (MP). The dimensions are tabulated in table \ref{tab:Detector parameters}. The FEP efficiency of each $\gamma$-ray was determined using the simulated spectrum and is shown in figure \ref{fig:monoenergetic} for add-back (left panel) and direct modes (right panel). From the figure it is observed that the simulated efficiencies are overestimated compared to the experimental efficiencies for the add-back and direct modes. Similar observations were also reported earlier in the literature \cite{gupta2023calculation, agarwal2011full, budjavs2009optimisation, vargas2002influence}. This overestimation in the simulated efficiency may be due to the inaccurate structural dimensions, such as crystal radius, crystal length, dead layer thickness, Al end cap to crystal distance ($d_{alc}$), etc. The $d_{alc}$ is very sensitive for close geometry measurements since a small change in this parameter may induce a significant change in the efficiency of the detector \cite{agarwal2014coincidence, agarwal2011full}. As the detector dimensions may be inaccurate, one needs to optimize the parameters using the experimentally measured efficiency data of mono-energetic $\gamma$-ray sources. The simulated efficiencies were matched to the experimental efficiencies by systematic optimization of the detector parameters. All the parameters of the detectors were tuned and optimized within the practical limitations. The simulated efficiencies with the optimized (O) detector parameters are shown in figure \ref{fig:monoenergetic} for the add-back mode and direct mode. From the figure, we can say that the experimental efficiencies corresponding to the $\gamma$-ray energies from the mono-energetic $\gamma$-ray sources match well with the simulated efficiencies within the error bars. The optimized detector dimension parameters are tabulated in table \ref{tab:Detector parameters}. The experimental efficiency and the simulated efficiency using the dimensions provided by the manufacturer as well as the parameters obtained after optimization are compared in table \ref{tab:monoenergetic_optimization}. The experimental efficiencies are found to be well reproduced by the Geant4 simulation done using the optimized set of parameters. The optimized detector dimension parameters can thus be used to determine the detector efficiency for any unknown $\gamma$-ray of interest.

\section{Results and discussion}
\label{Discussion}
The optimized detector parameters were used to simulate the $\gamma$-ray FEP corresponding to $^{60}$Co, $^{133}$Ba and $^{152}$Eu multi-energetic $\gamma$-ray sources. From the simulated spectrum, FEP and total efficiencies of the corresponding $\gamma$-rays were calculated for the distances $d$ = 25, 13, 5, and 1 cm. The FEP efficiencies calculated from the simulated spectrum are free from the coincidence summing effect. Hence the ratio of the FEP efficiency obtained from the simulated spectrum to the FEP efficiency obtained experimentally from the multi-energetic $\gamma$-ray source may be referred to as the experimental correction factor ($k^{Exp}_{TCS}$), as given by eq. \ref{eq:exp_correction_factor}. The $k^{Exp}_{TCS}$ for each distance has been determined for add-back and direct modes and is given in table \ref{tab:correctionfactor}.

\begin{figure}
    \centering
    \includegraphics[trim={1.8cm 1.5cm 2cm 1cm},clip,scale=0.6]{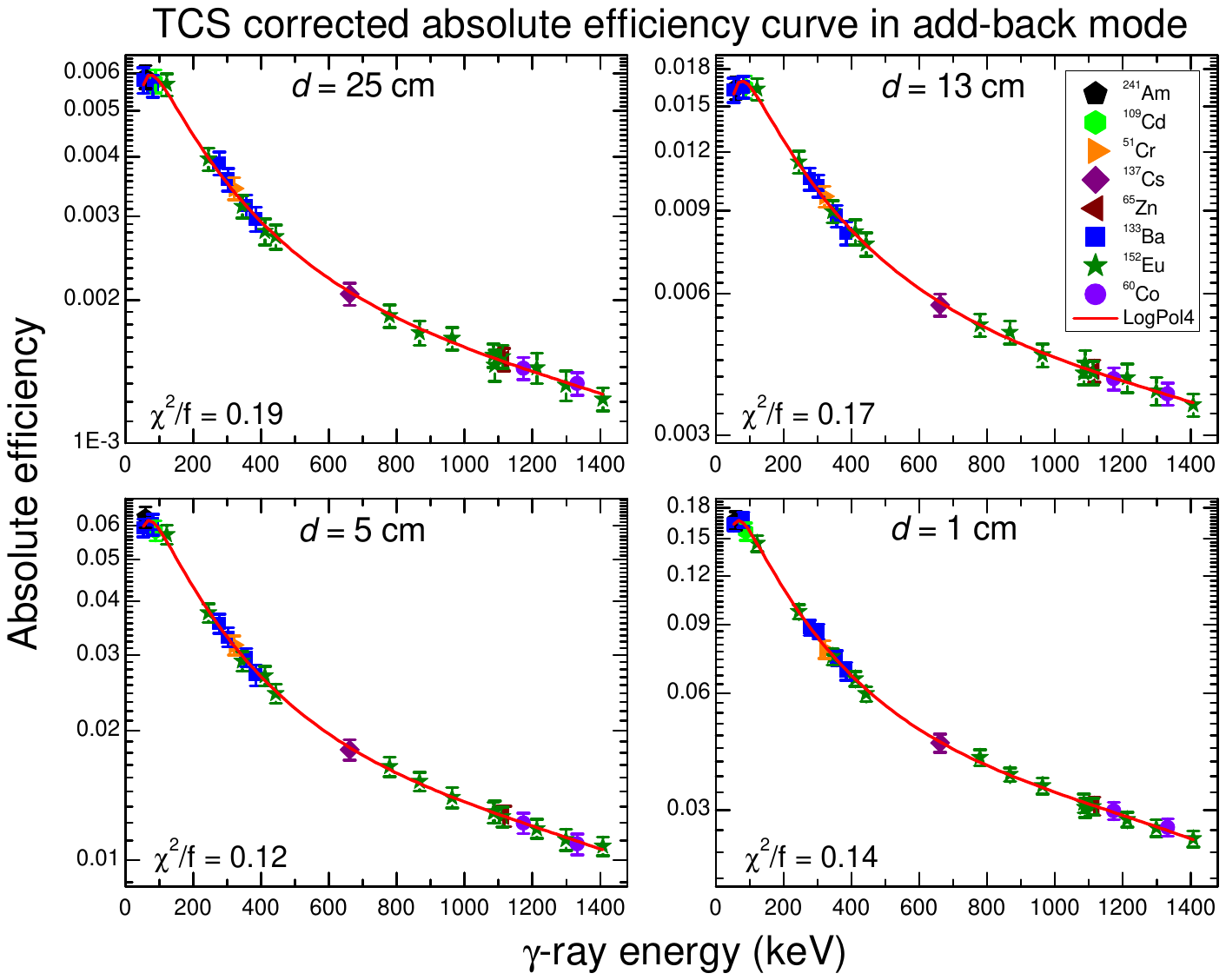}
    \caption{TCS corrected experimental efficiency for the add-back mode of clover detector as a function of $\gamma$-ray energy at different source-to-detector distances, $d$ = 25, 13, 5 and 1 cm. The solid lines show the fitted curves using the $4^{th}$ order log-log polynomial.}
    \label{fig:exp_ana_corected_addback_eff}
\end{figure}

The coincidence summing correction factors for the $\gamma$-rays of multi-energetic sources, $^{60}$Co, $^{133}$Ba, $^{152}$Eu were also calculated using the analytical method at each distance. To calculate the correction factor using the analytical method, one requires the probability of a particular $\gamma$-ray of interest being in coincidence with the other cascading $\gamma$-ray and also the total efficiency of the corresponding cascading $\gamma$-ray. The calculation of the probability requires information on the decay schemes and conversion coefficients. These were taken from the literature \cite{Browne1978,recomendeddata}. The FEP ($\epsilon_{i}$) and total efficiency ($\epsilon_{ti}$) of the $\gamma$-ray energies were simulated using the Geant4 Monte Carlo simulation toolkit. The analytical true coincidence summing correction factors ($k^{Analytical}_{TCS}$) obtained using eq. \ref{eq:analytical_correction_factor} are also tabulated in the table \ref{tab:correctionfactor}. The correction factors obtained from both methods were in good agreement with each other, for both add-back and direct modes.

\begin{figure}
    \centering
    \includegraphics[trim={1.5cm 1.2cm 2cm 1cm},clip,scale=0.6]{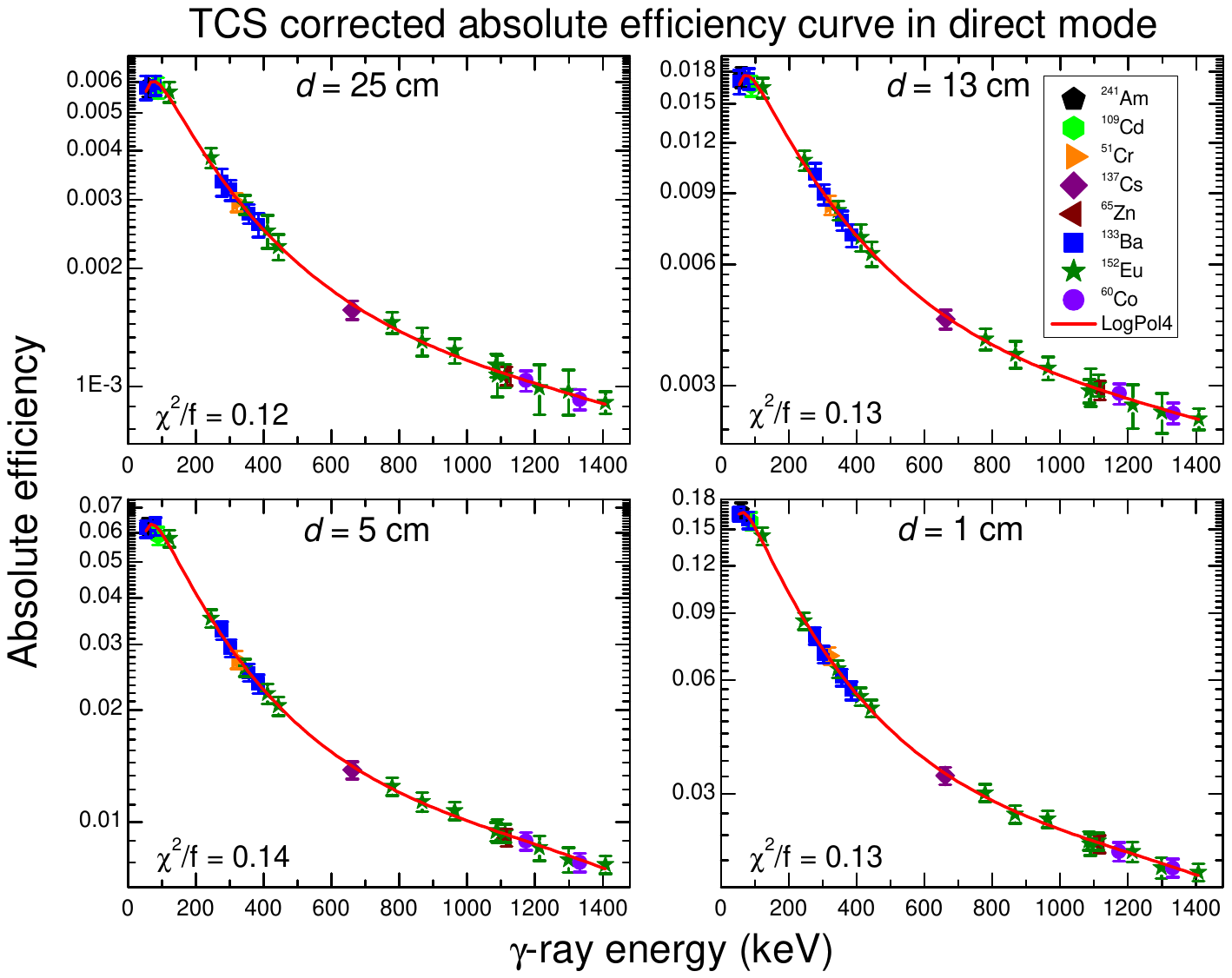}
    \caption{TCS corrected experimental efficiency for the direct mode of clover detector as a function of $\gamma$-ray energy at different source-to-detector distances, $d$ = 25, 13, 5 and 1 cm. The solid lines show the fitted curves using the $4^{th}$ order log-log polynomial.}
    \label{fig:exp_ana_corrected_sum_eff}
\end{figure}

The $k^{Analytical}_{TCS}$ and $k^{Exp}_{TCS}$ are compared at each distance for add-back and direct mode and plotted in figure \ref{fig:correction_factor}. The left panel shows the comparison of the correction factors between add-back and direct mode obtained by the experimental method whereas the right panel shows the same comparison for the analytical method. From the figure, it can be clearly seen that the correction factors for the add-back mode are higher than the direct mode. The correction factors for the add-back mode lie close to the unity for distances $\gtrsim$ 13 cm, for both analytical and experimental methods. Whereas, for the close geometry setup, it deviates from the unity for $d <$ 13 cm, and the deviation is quite large at $d$ = 1 cm. The correction factors for the direct mode lie close to unity for distances $\gtrsim$ 5 cm for both analytical and experimental methods. At $d$ = 1 cm correction factor deviates from the unity but is still significantly small compared to the correction factors in the add-back mode. For the add-back and direct mode of the clover detector, one needs to perform the summing correction if the measurements are taken at $d \lesssim$ 13 cm and $d \lesssim $ 5 cm, respectively.

The figures \ref{fig:exp_ana_corected_addback_eff} and \ref{fig:exp_ana_corrected_sum_eff} represent the experimental efficiency corrected for the coincidence summing correction factor obtained by the analytical method for the add-back mode and direct mode, respectively. The sum corrected experimental efficiency data points were fitted with a log-log polynomial of $4^{th}$ order. The value of $\chi^2/f$ given in each of the figures shows the quality of fit. The experimental data points after the summing corrections match well with the fitting curve. 

\section{Summary}
\label{summary}
The experimental full-energy peak efficiency of the clover detector in add-back and direct mode has been measured using the standard as well as fabricated mono-energetic and multi-energetic $\gamma$-ray sources for close and distant geometry. The true coincidence summing correction factor for the $\gamma$-ray energies emitted by multi-energetic $\gamma$-ray sources  $^{60}$Co, $^{133}$Ba, and $^{152}$Eu for the detector in both add-back and direct modes have been calculated using both analytical and experimental methods. Both experimental and analytical methods were found to be in good agreement with each other. The detector response of the clover detector was simulated using the Geant4 Monte Carlo simulation toolkit. The full-energy peak and total efficiencies of the detector were obtained from the simulated spectra. The simulated efficiencies were used in the calculation of the correction factor using the analytical method. It was found that the manufacturer provided detector dimension parameters had to be optimized to obtain accurate efficiencies. The results of this work show that for a clover detector coincidence summing correction is insignificant for the distances $\gtrsim$ 13 cm in the add-back mode and for distances $\gtrsim$ 5 cm in the direct mode. It is also seen that in a close geometry set-up, for the same source-to-detector distances, the correction factors are larger in the add-back mode compared to the direct mode for the clover detector. The clover detectors are commonly used in nuclear physics experiments and the detector is mostly used in the add-back mode for an increased detection efficiency of high energy $\gamma$-rays. However, in close geometry measurements, the add-back mode exhibits a larger summing effect compared to the direct mode. In the close geometry setup with a clover detector, the direct mode is an inevitable choice to perform the measurements rather than the add-back mode.


\bibliographystyle{JHEP} 
\bibliography{ref}
 






\end{document}